\documentclass[11pt,a4paper,dvipdfm]{article}
\usepackage{graphicx}

\usepackage[ruled,lined,algonl,boxed]{algorithm2e}

\begin{document}
\newcommand{\up}{\vspace*{-0.05cm}}
\newcommand{\qed}{\hfill$\rule{.05in}{.1in}$\vspace{.3cm}}
\newcommand{\pf}{\noindent{\bf Proof: }}
\newtheorem{thm}{Theorem}
\newtheorem{lem}{Lemma}
\newtheorem{prop}{Proposition}
\newtheorem{prob}{Problem}
\newtheorem{quest}{Question}
\newtheorem{ex}{Example}
\newtheorem{cor}{Corollary}
\newtheorem{conj}{Conjecture}
\newtheorem{cl}{Claim}
\newtheorem{df}{Definition}
\newtheorem{rem}{Remark}
\newcommand{\beq}{\begin{equation}}
\newcommand{\eeq}{\end{equation}}
\newcommand{\<}[1]{\left\langle{#1}\right\rangle}
\newcommand{\be}{\beta}
\newcommand{\ee}{\end{enumerate}}
\newcommand{\Bul}{\mbox{$\bullet$ } }
\newcommand{\al}{\alpha}
\newcommand{\ep}{\epsilon}
\newcommand{\si}{\sigma}
\newcommand{\om}{\omega}
\newcommand{\la}{\lambda}
\newcommand{\La}{\Lambda}
\newcommand{\Ga}{\Gamma}
\newcommand{\ga}{\gamma}
\newcommand{\im}{\Rightarrow}
\newcommand{\2}{\vspace{.2cm}}
\newcommand{\es}{\emptyset}

\vspace{-2cm}

\title{\LARGE\bf Bounds and algorithms for limited packings in graphs\footnote{extended abstract, appeared in the Proceedings of the 9th International Colloquium on Graph Theory and Combinatorics, ICGT 2014, Grenoble, France, June 30\,-\,July 4, 2014, paper no.\,27}}
\author{Andrei Gagarin\footnote{e-mail: {\tt andrei.gagarin@rhul.ac.uk}}\\ 
{\footnotesize Royal Holloway, University of London, Egham, Surrey, TW20 0EX, UK}\vspace{4mm}\\
Vadim Zverovich\footnote{e-mail: {\tt vadim.zverovich@uwe.ac.uk}}\\
{\footnotesize University of the West of England, Bristol, BS16 1QY, UK}
}
\date{}
\maketitle
\begin{abstract}
We consider (closed neighbourhood) packings and their generalization in graphs called limited packings.
A vertex set $X$ in a graph $G$ is a {\it $k$-limited packing} if for any vertex $v\in V(G)$,
$\left|N[v] \cap X\right| \le k$, 
where $N[v]$ is the closed neighbourhood of $v$. 
The {\it $k$-limited packing number} $L_k(G)$ is the largest size of 
a $k$-limited packing in $G$. Limited packing problems can be considered as secure facility location problems in networks.
We develop probabilistic and greedy approaches to limited packings in graphs, providing lower bounds for the $k$-limited packing number, and randomized and greedy algorithms to find $k$-limited packings satisfying the bounds. 
Some upper bounds for $L_k(G)$ are given as well. The problem of finding a maximum size $k$-limited packing is known to be $NP$-complete even in split or bipartite graphs.
\end{abstract}

\medskip
\section{Introduction}
\label{intro}
\noindent We consider simple undirected graphs, and are interested in the classical packings of graphs as introduced in \cite{MM75}, and their generalization, called limited packings, as presented in \cite{GGHR10}. In the literature, the classical packings can be referred to under different names, e.g., as (distance) $2$-packings \cite{MM75}, closed neighborhood packings \cite{RSS06}, or strong stable sets \cite{HS85}. They can also be considered as generalizations of independent (stable) sets, which would be (distance) $1$-packings, following the terminology of \cite{MM75}. 
A vertex set $X$ in a graph $G$ is a {\it $k$-limited packing} if for any vertex $v\in V(G)$,
$\left|N[v] \cap X\right| \le k$, 
where $N[v]$ is the closed neighbourhood of $v$. 
The {\it $k$-limited packing number} $L_k(G)$ of a graph $G$ is the maximum size of 
a $k$-limited packing in $G$. In these terms, the classical (distance) $2$-packings are $1$-limited packings, and hence $\rho(G)=L_1(G)$, where $\rho(G)$ is the 2-packing number.

$2$-Packings ($1$-limited packings) are well-studied in the literature (e.g., \cite{HLR11,HS85,MM75,RSS06}). Also, several papers discuss connections between packings and dominating sets in graphs (e.g., \cite{DLN11, GGHR10, HLR11, RSS06}). Notice that the corresponding problems have a very different nature: one of the problems (packings) is a maximization problem not to break some (security) constraints, and the other is a minimization problem (dominating sets) to satisfy some reliability requirements. For example, given a graph $G$, the definitions imply a simple inequality $\rho(G)\le\gamma(G)$, where $\gamma(G)$ is the domination number of $G$ (e.g., see \cite{RSS06}). However, the difference between $\rho(G)$ and $\gamma(G)$ can be arbitrarily large.

In general, the problems of packings in graphs can be viewed as facility location problems subject to some (security) constraints.
The problem of finding a $2$-packing ($1$-limited packing) of maximum size is shown to be $NP$-complete in \cite{HS85}. In \cite{DLN11}, it is shown that the problem of finding a maximum size $k$-limited packing is $NP$-complete even in split or bipartite graphs.

We develop the probabilistic method 
for $k$-limited packings and, in particular, for $2$-packings ($1$-limited packings), with a comparison to a greedy approach. 
We present two new lower bounds for the $k$-limited packing number $L_k(G)$, and a randomized algorithm to find $k$-limited packings satisfying the lower bounds. 
Also, using a greedy algorithm approach, we provide an improved lower bound for $\rho(G)=L_1(G)$. 
It can be shown that the main lower bound is asymptotically sharp, and the improvements for $1$-limited packings from the greedy algorithm approach are asymptotically irrelevant for almost all graphs.
We also provide some new upper bounds for $L_k(G)$.
Details for some of these results can be found in \cite{GZ2012}.

\section{Lower bounds and algorithms}
\label{main}
Denote by $\Delta$ = $\Delta(G)$ the maximum vertex degree of $G$, and put $n=|V(G)|$. Notice that $L_k(G)=n$ for $k\ge\Delta+1$.
We put
${\tilde c}_t = {\tilde c}_t (G) =  {\Delta+1 \choose t}$.
The following theorem gives a new lower bound for the $k$-limited packing number.

\begin{thm} \label{th1}
For any graph $G$ of order $n$ with $\Delta\ge k \ge 1$,
\begin{equation} \label{main_bound}
L_k(G) \ge {k n \over {\tilde c}_{k+1}
^{1/k} \; (1+k)^{1+1/k}}.
\end{equation}
\end{thm}

The probabilistic construction used to prove Theorem \ref{th1}  implies a randomized algorithm to find a $k$-limited packing  set, whose size satisfies bound (\ref{main_bound}) with a positive
probability.
A pseudocode presented in Algorithm \ref{alg:k-limited} below explicitly describes the randomized algorithm.
Algorithm \ref{alg:k-limited} can be implemented to run in $O(n^2)$ time. 
Also, it can be implemented in parallel or as a local distributed algorithm, which is important in some applications.

\begin{algorithm}\label{alg:k-limited}
    \caption{Randomized $k$-limited packing}

    \KwIn{Graph $G$ and integer $k$, $1\le k\le
\Delta$.}
    \KwOut{$k$-Limited packing $X$ in $G$.}
    \BlankLine
\Begin{

Compute $p = \left({1 \over {{\Delta \choose k} \cdot (\Delta+1)}}\right)^{1/k}\ $\;
\SetLine
Initialize $A=\emptyset$;\tcc*[f]{Form a set $A\subseteq V(G)$}\\
\ForEach{ vertex $v\in V(G)$} {
    with the probability $p$, decide whether $v\in A$ or
$v\notin A$\; }
\SetLine
\tcc*[f]{Recursively remove redundant vertices from $A$}\\
\ForEach{ vertex $v\in V(G)$} {
 Compute $r=|N(v)\cap A|$\;
    \If{$v\in A$ and $r\ge k$}
        {remove any $r-k+1$ vertices of $N(v)\cap A$ from $A$\;
    }
    \If{$v\notin A$ and $r>k$}
    	{remove any $r-k$ vertices of $N(v)\cap A$ from $A$\;
    }
}
Put $X = A$;\tcc*[f]{$A$ is a $k$-limited packing}\\
Extend $X$ to a maximal $k$-limited packing\;
\Return $X$;
}
\end{algorithm}

The lower bound of Theorem \ref{th1} can be written in a simple weaker form as follows:

\begin{cor}
For any graph $G$ of order $n$,
$$
L_k(G) > {kn \over e (1+\Delta)^{1+1/k}}.
$$
\end{cor}

\noindent In the case $k=1$, Theorem \ref{th1} gives the following lower bound:
\begin{cor} \label{cor22}
For any graph $G$ of order $n$ with $\Delta\ge 1$,
\begin{equation}
\label{bcor22}
\rho(G)=L_1(G) \ge {n \over 2\Delta(\Delta +1)}.
\end{equation}
\end{cor}

Let $\delta$ = $\delta(G)$ denote the minimum vertex degree of $G$. The lower bound (\ref{bcor22}) can be improved as follows:

\begin{thm} \label{th22}
For any graph $G$ of order $n$,
\begin{equation} \label{greedy_bound}
\rho(G)=L_1(G) \ge {n+\Delta(\Delta-\delta) \over \Delta^2 +1} \ge {n \over \Delta^2 +1}.
\end{equation}
\end{thm}

\noindent The proof of Theorem \ref{th22} provides a greedy algorithm to find a distance $2$-packing ($1$-limited packing) satisfying bound (\ref{greedy_bound}). However, as explained later, the lower bound of Theorem \ref{th22} is as good as lower bound (\ref{bcor22}) of Corollary \ref{cor22} for almost all graphs.

Refining and developing the ideas and techniques used in the proofs of the above results, we have been able to show the following more subtle results.

\begin{thm} \label{thA}
For any graph $G$ of order $n$, with vertex degrees $d_i=\mathrm{deg}(v_i)$, $i=1,\ldots,n$,
\begin{equation} \label{thA_bound}
\rho(G)=L_1(G) \ge\sum_{i=1}^{n} {1 \over d_i\cdot\Delta +1}.
\end{equation}
\end{thm}

\begin{thm} \label{thB}
For any $\Delta$-regular graph $G$ of order $n$, $\Delta\ge 3$,
\begin{equation} \label{thB_bound}
L_2(G) \ge {2n \over \Delta^2 - \Delta +2}.
\end{equation}
\end{thm}


\section{Sharpness of the lower bounds and some upper bounds}

The lower bound of Theorem \ref{th1} is asymptotically best possible for some `large' values of $k$. Bound (\ref{main_bound}) can be written as
$L_k(G) \ge {k n \over (k+1) \sqrt[k]{{\Delta \choose k} (\Delta +1)} }$ for $k \le \Delta$.
Combining this with the upper bound of Lemma $8$ from \cite{GGHR10},
for any connected graph $G$ with $k \le \delta(G)$,
we have:
\begin{equation} \label{ineq5}
{1 \over \sqrt[k]{{\Delta \choose k} (\Delta +1)} }\times {k \over k+1} n  \;\le\;
L_k(G)\; \le \;{k \over k+1} n.
\end{equation}
Notice that the upper bound in the inequality (\ref{ineq5}) is sharp (see \cite{GGHR10}), and bounds (\ref{ineq5}) provide an interval of possible values for $L_k(G)$ in terms of $k$ and $\Delta$ (when $k\le\delta$). For regular graphs, $\delta=\Delta$, and, when $k=\Delta$, we have
${1 \over \sqrt[k]{{\Delta \choose k} (\Delta +1)} } = 
{1 \over \left(k+1\right)^{1/k}} \longrightarrow 1\ \quad\mbox{as} \quad \ k \to \infty$.
Therefore, Theorem \ref{th1} is asymptotically sharp for regular connected graphs in the case $k=\Delta$. 
A similar statement can be proved when $k=\Delta (1-o(1))$. 
Thus, the following result is true:

\begin{thm} \label{asym_sharp}
When $n$ is large, there exist graphs $G$ such that 
\begin{equation} 
L_k(G) \le {k n \over {\tilde c}_{k+1}
^{1/k} \; (1+k)^{1+1/k}}(1+o(1)).
\end{equation}
\end{thm}

As stated in Theorem \ref{th22}, in contrast to the situation for relatively `large' values of $k$, bound (\ref{main_bound}) of Theorem \ref{th1} (Corollary \ref{cor22}) can be improved for distance $2$-packings ($1$-limited packings), i.e. when $k=1$. However, this improvement is irrelevant for almost all graphs. A $1$-limited packing set $X$ in $G$ has a very strong property that any two vertices of $X$ are at distance at least $3$ in $G$. It is well known that almost every graph has diameter equal to $2$. Therefore, $\rho(G)=L_1(G)=1$ for almost all graphs. Thus, in the case $k=1$, Theorem \ref{th1} yields a lower bound of 1 for almost all graphs and is as good as Theorem \ref{th22}. Notice that the bound of Theorem \ref{th22} is sharp, for example for any number of disjoint copies of the Petersen graph. When $G$ has a diameter larger than $2$, it should be better to use the greedy algorithm and lower bound (\ref{greedy_bound}) provided by Theorem \ref{th22}: it improves bound (\ref{bcor22}) of Corollary \ref{cor22} by a factor of $2+o(1)$.
 

As mentioned earlier, $\rho(G)=L_1(G)\le\gamma(G)$. In \cite{GGHR10}, the authors provide several upper bounds for $L_k(G)$, e.g., given any graph $G$, $L_k(G)\le k\gamma(G)$.
Considering the $k$-tuple domination number $\gamma_{\times k}(G)$ and results of \cite{GPZ13}, it is possible to prove the following upper bounds for $L_k(G)$. 
A set $X$ is a \emph{$k$-tuple dominating set} of $G$ if for every vertex $v\in V(G)$,
$|N[v]\cap X|\ge k$. The minimum cardinality of a $k$-tuple
dominating set of $G$ is the {\it $k$-tuple domination number}
$\gamma_{\times k}(G)$. The $k$-tuple domination number is only
defined for graphs with $\delta\ge k-1$.
In \cite{GPZ13}, we define
$\delta' = \delta -k +1$ 
and, for $t\le \delta$,
${\tilde b}_t = {\tilde b}_t (G) =  {\delta+1 \choose t}$.

\begin{thm} \label{th2}
For any graph $G$ of order $n$ with $\delta\ge k-1$,
$L_k(G)\le\gamma_{\times k}(G)$. 
In particular, when $k\le \delta$,
$$\qquad L_k(G)\le  \left(1-{\delta' \over {\tilde b}_{k-1}
^{1/\delta'} (1+\delta')^{1+1/\delta'}}\right) n.$$
\end{thm}

For regular graphs, we also have:

\begin{prop} \label{propositon_regular}
If $G$ is a $\Delta$-regular graph of order $n$, then\ \ $L_k(G)\le \frac{kn}{\Delta +1}$.
\end{prop}


\end{document}